\documentclass[twoside,slac_one]{revtex4}
\usepackage{graphicx}
\usepackage{fancyhdr}
\usepackage{amsmath} 
\usepackage{bm}
\usepackage{amsxtra}
\usepackage{amssymb}
\usepackage{amsthm}
\usepackage{latexsym}
\usepackage{lscape}

\newcommand{\ptmax}{p_T^{\rm{max}}}
\newcommand{\ptmin}{p_T^{\rm{min}}}
\newcommand{\Mjj}{M_{jj}}

\newcommand{\chijj}{\chi_{\rm{dijet}}}
\newcommand{\ystar}{y^*}

\newcommand{\as}{\alpha_s}

\newcommand{\ascb}{{\alpha_s^3}}

\newcommand{\Rcone}{R_{\rm{cone}}}
\newcommand{\Rtt}{R_{\rm{3/2}}}

\newcommand{\ppbar}{p{\bar{p}}}

\begin{document}

\title{Multijet Measurements with the D\O\ Detector}

%

\author{L. Sawyer}
\affiliation{Center for Applied Physics Studies, Louisiana Tech University, Ruston, LA, USA}

\begin{abstract}
In this note we present several recent results for the production of multijet final states in $p\bar{p}$ collisions at a center of mass energy of 1.96~TeV. These measurements, which include the cross-section for dijet and three-jet production and the ratio of three-jet to two-jet cross sections, were taken with the D\O\ experiment at the Fermilab Tevatron collider.
\end{abstract}

\maketitle

\thispagestyle{fancy}


\section{Introduction}
In hadron-hadron collisions, production rates of collimated sprays of hadrons,
called jets, are sensitive to both the dynamics of the fundamental interaction and to the
partonic structure of the initial-state hadrons. The latter is usually parameterized in parton distribution
functions (PDFs) of the hadrons. Thus measurements of jet production properties can be used to test the predictions of quantum chromodynamics (QCD) as well as to constrain the PDFs. Because jet production has the largest cross-section of any high $p_T$ process at a hadron collider, the study of jet production provides the highest energy reach for the experiment and a unique sensitivity to new physics.

Several jet measurements are now well established at the Tevatron, including the measurement of the inclusive jet cross section and of the angular and invariant mass dependencies of dijet events~\cite{2008hua,2009mh,2010dijm}. These measurements provide important tests of perturbative QCD (pQCD) and are input to PDF calculations. Measurements of multi-jet production take advantage of the fact that these processes have the same PDF sensitivity as dijet production, but are sensitive to processes to third order in the strong coupling constant $\as$. Fundamental to the understanding of order $\ascb$ processes is the three-jet cross-section, which we present as a function of the invariant mass of the three-jet system. Studies dedicated to the dynamics of the interaction are preferably based on observables which are insensitive to the PDFs. Such observables can be constructed as ratios of cross sections for which the PDF sensitivity cancels. In this note we report a measurement of $\Rtt$, the ratio of the inclusive three-jet to the inclusive 2-jet cross-sections.

In this note we present several recent results for the production of multijet final states in $\ppbar$ collisions at a center of mass energy of 1.96~TeV. These measurements, which include the cross-section for dijet and three-jet production and the ratio of three-jet to two-jet cross sections, where taken with the D\O\ experiment~\cite{d0det} at the Fermilab Tevatron collider. The data analyzed were recorded using a single jet triggers at a variety of thresholds, corresponding to an integrated luminosity of 0.7~fb$-1$.
The event selection, jet reconstruction, jet energy and momentum correction in this measurement follow closely those used in our recent measurements of inclusive jet and dijet distributions~\cite{2008hua,2009mh,2010dijm}. The primary tool for jet detection is the finely segmented uranium-liquid argon calorimeter that has almost complete solid angle coverage
$1.7^\circ < \theta < 178.3^\circ$~\cite{d0det}.
Jets are defined by the Run~II midpoint cone jet algorithm~\cite{run2cone} with a cone radius (for most jet studies) of
$\Rcone =\sqrt{(\Delta y)^2 + (\Delta \phi)^2}=0.7$ in rapidity $y$ and azimuthal angle $\phi$.
Rapidity is related to the polar scattering angle $\theta$ with respect to the beam axis
by $y=0.5 \ln \left[ (1+\beta \cos \theta) / (1-\beta \cos \theta) \right]$
with $\beta=|\vec{p}| / E$. The jets in an event are ordered in descending transverse momentum $p_T$
with respect to the beam axis.

For all jet studies, a jet energy scale (JES) correction is applied to jet four momentum as the measured four momentum of a jet is
not the same as that of a jet entering the calorimeter due to the response of the calorimeter; energy showering in and out of the cone; and additional energy from detector noise, event pile-up and multiple $\ppbar$ interactions. The JES correction is determined using the $p_T$ imbalance in $\gamma +$ jet and dijet events. The additional energy from pile-up and multiple interaction is determined from a minimum bias sample. The JES corrections are of the order of 50\% for a jet energy of 50 GeV and 20\% for a jet energy of 400 GeV. However, for the 0.7 pb$-1$ data sample used for the measurements presented in this talk, collected with the D\O\ detector during 2004--2005
in Run~II of the Fermilab Tevatron Collider, the uncertainty on the JES is very small, less than 2\% over jet $p_T$'s from 60 to 300 GeV.
\begin{figure}[ht]
\begin{centering}
\includegraphics[scale=0.60]{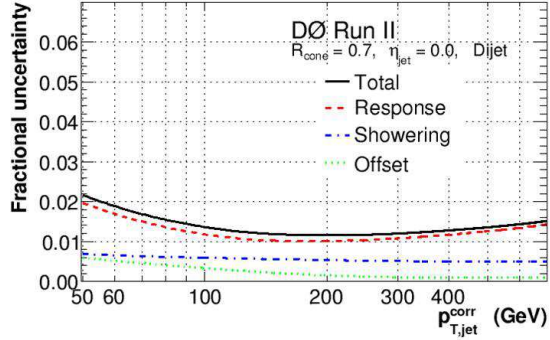}
\caption{\label{fig:JESfracUnc} The fractional uncertainty from the jet energy scale (JES) for jets of cone size $\Rcone = 0.7$ and central pseudorapidity.}
\end{centering}
\end{figure}

For the sake of brevity, the reader is directed to reference~\cite{2008hua} for a discussion of the inclusive jet cross section measurement, and reference~\cite{2009mw} for the details on the extraction of the strong coupling constant from the inclusive jet cross section.

\section{Dijet Cross-sections}

Of the variables that fully specify the kinematics of dijet production in hadron collisions, the dijet invariant mass $\Mjj$ and rapidity difference $\ystar$ (or equivalently $\chijj$) probe the features of LO $2\rightarrow 2$ processes, while measurements of the azimuthal opening angle $\Delta\phi$ and ratio of transverse momenta $p_{T2}/p_{T1}$ can be used to determine the effect of higher-order processes. The dijet invariant mass cross section is particularly sensitive to the gluon contribution to PDFs at high $x$. In addition, the dijet mass spectrum can be used to search for evidence of quark compositeness or new strongly produced resonances.

D\O\ has measured the dijet cross section as a function of the dijet invariant mass $\Mjj$\cite{2010dijm}. This cross section is measured in six regions of $y_{max}$, the rapidity of the leading jet in the event.

\begin{figure}[ht]
\begin{centering}
\includegraphics[scale=0.40]{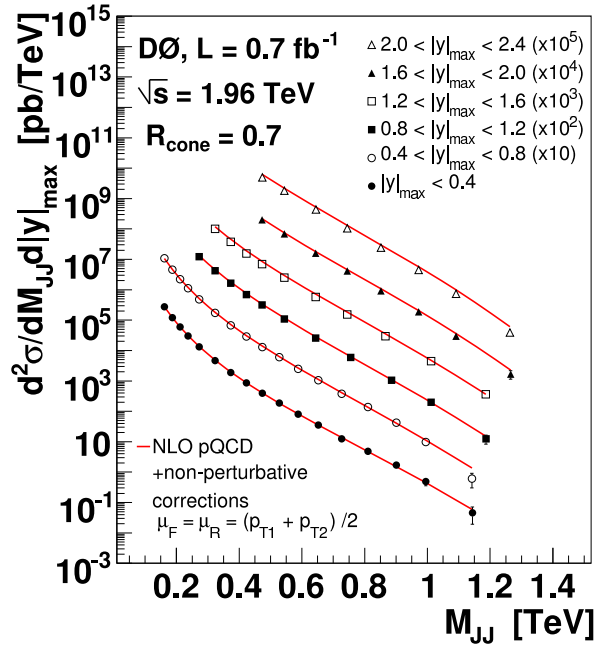}
\caption{\label{fig:dijetMassSpectra} Dijet mass cross sections, shown for six regions of $|y_{max}|$. (Note that these regions scaled for plotting, wtih the exception of $|y_{max}|<0.4$.) Full lines correspond to the NLO calculations with NLOJET++ and MSTW2008 PDFs.}
\end{centering}
\end{figure}

Figure~\ref{fig:dijetMassSpectra} shows the resulting measurements, while figure~\ref{fig:dijetMassSpectra} shows the measured cross section divided by the NLO prediction from the {\tt NLOJET++}~\cite{nlojet} program and using the MSTW2008NLO PDFs~\cite{MSTW2008}. This measurement in an important extension of pQCD to the forward region, and includes measurements of dijet invariant masses up to 1.2~TeV.

\begin{figure}[ht]
\begin{centering}
\includegraphics[scale=0.60]{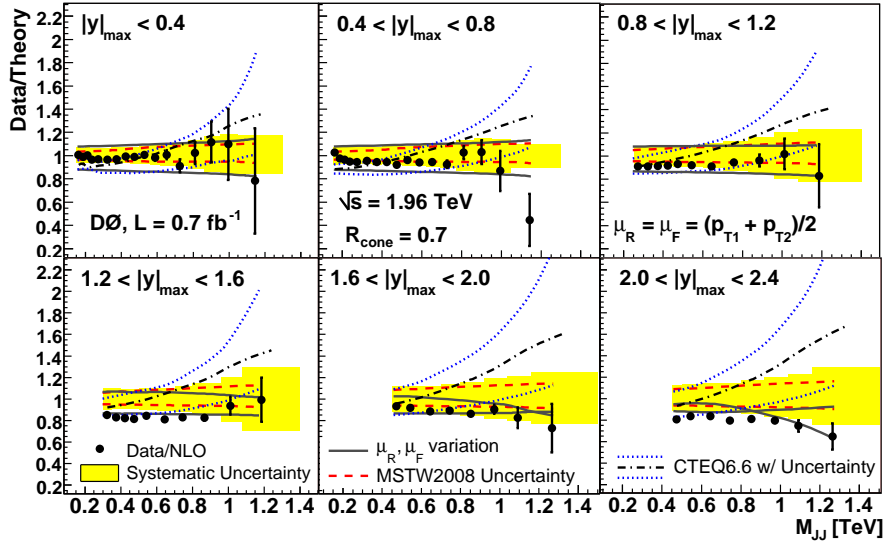}
\caption{\label{fig:dijetMassDataTheory} Dijet mass cross section divided by NLO prediction, with total systematic uncertainties indicated by the shaded bands.The dataset corresponds to an integrated luminosity of 0.7 fb$^{-1}$.}
\end{centering}
\end{figure}

\section{Three Jet Cross-section}

 D\O\ has measured~\cite{2011trijm} the differential inclusive three-jet cross section as a function of the invariant three-jet mass for three hard, well separated jets in three regions of jet rapidities ($|y| < 0.8$, $|y| < 1.6$, $|y| < 2.4$~for all three jets) and in three regions of the third jet transverse momentum ($p_{T3} > 40$~GeV, $p_{T3} > 70$~GeV and $p_{T3} > 100$~GeV) with leading jet $p_T > 150$~GeV. The three-jet mass spectra are shown in Fig.~\ref{fig:M3jvsY}. This represents the first fully corrected 3-jet mass distribution to be published, and the first comparison of such a cross-section to NLO pQCD calcuations.

\begin{figure}[ht]
\begin{centering}
\includegraphics[scale=0.90]{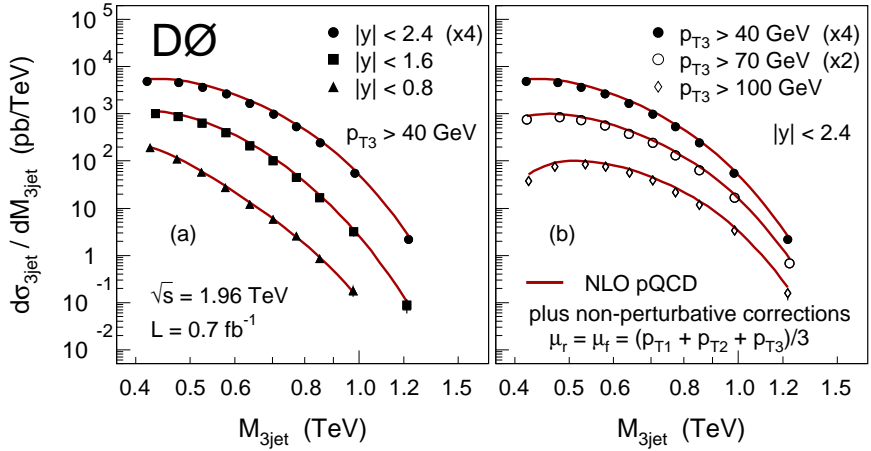}%
\caption{\label{fig:M3jvsYvspT3} a) Three-jet mass cross section in regions of jet rapidities. The $|y| < 2.4$ region is scaled by a factor of 4 for readability. Systematic uncertainty is shown by shaded band. Full lines correspond to the NLO calculations with NLOJET++ and MSTW2008 PDFs. No events are found in the highest $M_{\rm 3jet}$ bin in $|y| < 0.8$ region. b) Three-jet mass cross section in regions of the third jet transverse momenta. The $p_{T3} > 40$~GeV region is scaled by a factor of 2 for readability. Systematic uncertainty in all three-jet mass bins is shown by shaded band. Full lines correspond to the NLO calculations with NLOJET++ and MSTW2008 PDFs. In both figures, the dataset corresponds to an integrated luminosity of 0.7 fb$^{-1}$.}
\end{centering}
\end{figure}

\begin{figure}[ht]
\begin{centering}
\includegraphics[scale=0.80]{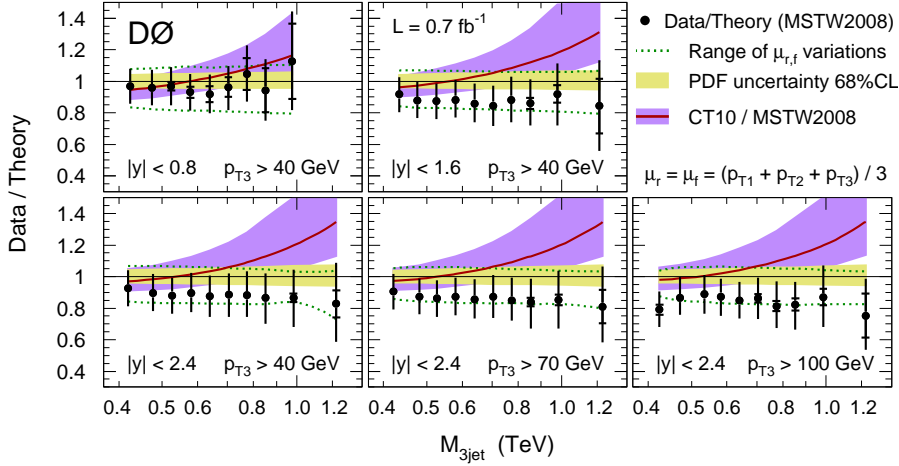}%
\caption{\label{fig:M3jDThvsY} Data to theory ratio of the three-jet mass cross section in three regions of jet rapidities.The total systematic uncertainty is shown by a shaded band. The PDF uncertainty comes from the 20 MSTW2008NLO eigenvectors. The scale uncertainty is determined by varying the scale up and down by a factor of 2.}
\end{centering}
\end{figure}

\begin{figure}[htp]
\begin{centering}
\includegraphics[scale=0.80]{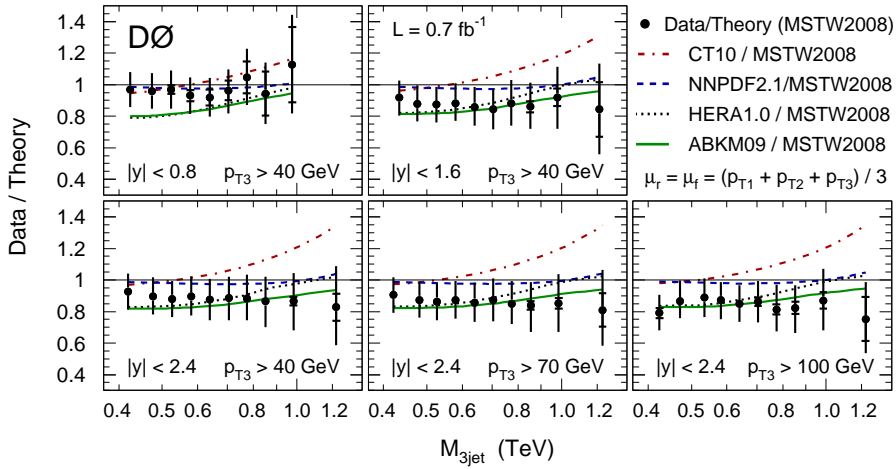}%
\caption{\label{fig:M3jDThvspT3} Data to theory ratio of the three-jet cross section in three regions of the third jet transverse momentum. The total systematic uncertainty is shown by a shaded band. The PDF uncertainty comes from the 20 MSTW2008NLO eigenvectors. The scale uncertainty is determined by varying the scale up and down by a factor of 2.}
\end{centering}
\end{figure}

The data are compared to the perturbative QCD NLO predictions calculated in NLOJET++\cite{nlojet} program with MSTW2008NLO\cite{MSTW2008} PDFs. The NLO prediction is corrected for hadronization and
underlying event effects, with correction varied in the range from -3\% to 6\%, obtained from PYTHIA\cite{pythia} simulations using tune QW\cite{tuneQW}. The comparison of data to theory is shown in Fig.~\ref{fig:M3jDThvsY} and~\ref{fig:M3jDThvspT3}.


\section{Ratio of Three-jet to Dijet Cross-section, $\Rtt$}

The ratio of inclusive jet cross sections,
$\Rtt(\ptmax) = (d\sigma_{\ge\rm{3jet}}/d\ptmax) / (d\sigma_{\ge\rm{2-jet}}/d\ptmax)$,
is less sensitive to experimental and theoretical uncertainties than the individual cross sections,
due to cancelations of correlated uncertainties. Here $\Rtt$ is measured as a function of the leading jet $p_T$ in an event, $\ptmax$, in the interval $(\ptmin + 30\,{\rm GeV}) < \ptmax < 500\,{\rm GeV}$,
for $\ptmin$ requirements of 50, 70, and 90 GeV.

\begin{figure}[ht]
\begin{centering}
\includegraphics[scale=0.80]{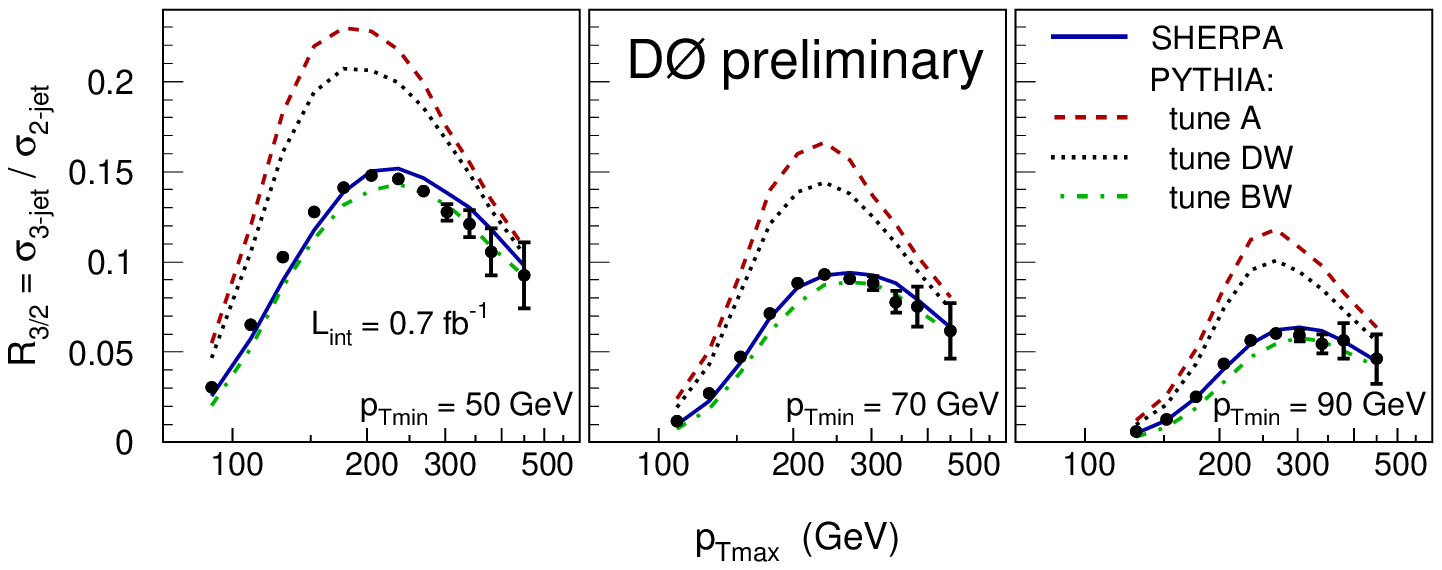}%

\includegraphics[scale=0.80]{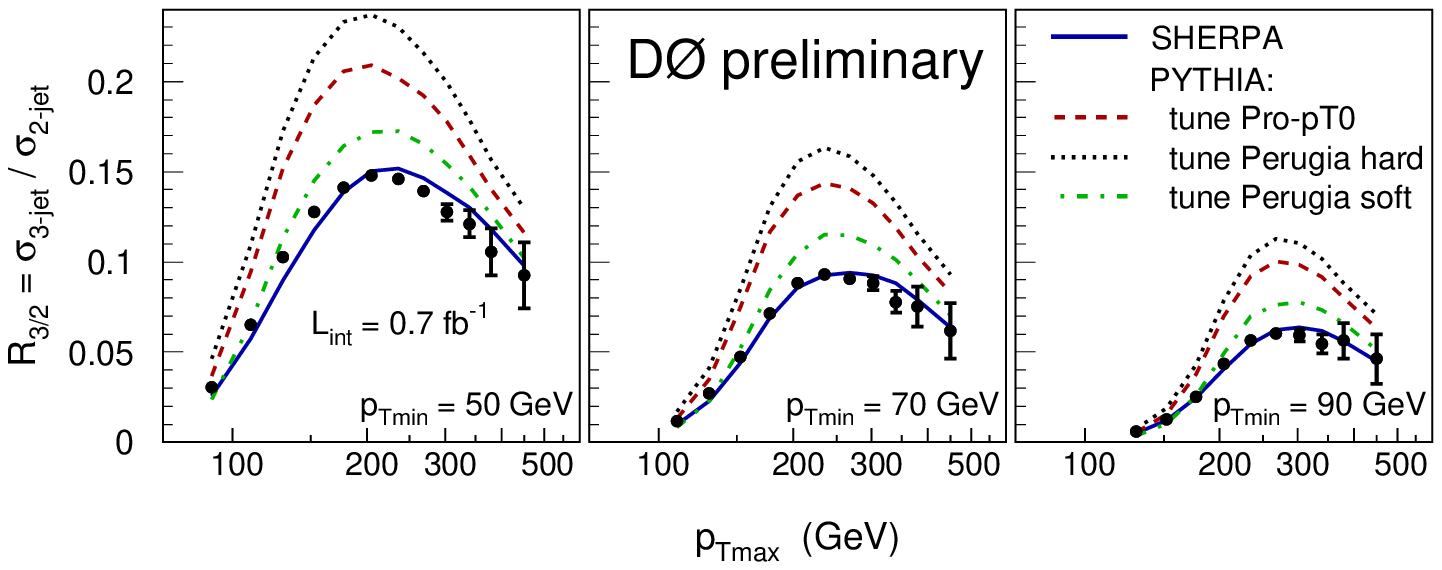}%
\caption{\label{fig:result}The ratio $\Rtt$ of trijet and dijet cross sections, measured as a function of the leading jet $p_T$ ($\ptmax$) for different $\ptmin$ requirements for the other jets. The predictions of {\sc sherpa} and {\sc pythia} (for Tunes A, BW, and DW (top); and for three tunes using the $p_T$-ordered parton shower (bottom)) are compared to the data.}
\end{centering}
\end{figure}

The results are displayed in Fig.~\ref{fig:result},
where the inner error bars represent the statistical uncertainties
while the total error bars represent the quadratic sums of statistical
and systematic uncertainties. The data are compared to the predictions from different Monte Carlo event generators.
The {\sc sherpa} v1.1.3 predictions~\cite{Gleisberg:2008ta}, which include the tree-level matrix elements for 2-, 3-, and 4-jet production, are shown as solid lines in Figs.~\ref{fig:result}. These were obtained using default settings and MSTW2008LO PDFs and by matching the leading order matrix elements for 2-, 3-, and 4-jet production with a parton shower.

In addition, the data are compared to predictions from the PYTHIA event generator (version 6.422). The matrix elements implemented in PYTHIA are only those for 2-jet production. All additional jet emissions are produced by a parton shower. There are two different implementations, a virtuality-ordered parton shower
and a $p_T$-ordered one. Both are highly tunable and more than 50 tunes are provided
in PYTHIA v6.422. All tunes studies here use the CTEQ5L PDFs~\cite{Lai:1999wy}.

In Fig.~\ref{fig:result} the data are compared to PYTHIA tunes
which use the $p_T$-ordered parton shower~\cite{Sjostrand:2004ef,Buttar:2006zd}.
These are tune ``Professor pT0''~\cite{Buckley:2009vk}
and two extreme tunes from the ``Perugia'' series of tunes~\cite{Skands:2009zm},
the tunes ``Perugia hard'' and ``Perugia soft''.
All of these tunes give very different results for $\Rtt$
but all predict significantly higher ratios $\Rtt$ than what
is seen in the data, even for the softest tune from the Perugia series.

\begin{figure}[ht]
\begin{centering}
\includegraphics[scale=0.80]{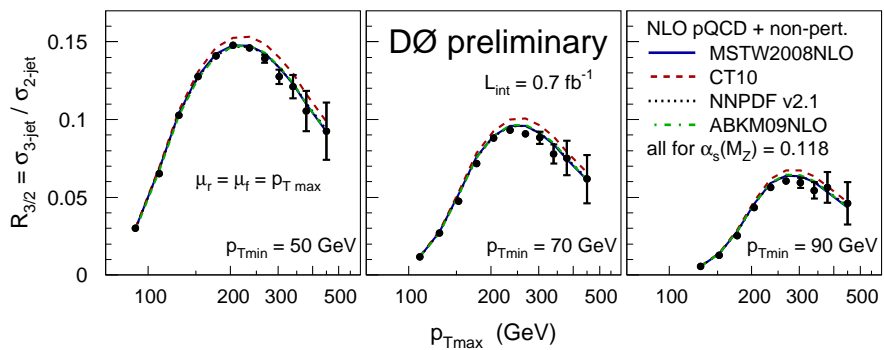}%
\caption{\label{fig:result2}
  The ratio $\Rtt$ of trijet and dijet cross sections, measured
  as a function of the leading jet $p_T$ ($\ptmax$) for different
  $\ptmin$ requirements for the other jets.
  The predictions of next-to-leading order pQCD, calculated with NLOJET++, are compared to the data.
  }
\end{centering}
\end{figure}

The data are compared to the perturbative QCD NLO predictions calculated in NLOJET++\cite{nlojet} program with MSTW2008NLO\cite{MSTW2008} PDFs. The NLO prediction is corrected for hadronization and
underlying event effects, with correction varied in the range from -3\% to 6\%, obtained from PYTHIA\cite{pythia} simulations using tune QW\cite{tuneQW}. The comparison of data to theory is shown in Fig.~\ref{fig:result2}.

\section{Summary}

The D\O\ experiment at the Fermilab Tevatron collider has been enormously successful, and continues to produce important physics measurements, including tests of the Standard Model and searches for new phenomena. Only a small portion of the QCD-related analyzes we have pursued could be shown in this talk. The reader is urged to check \url{http://www-d0.fnal.gov/results/index.html} for the latest results.

\begin{acknowledgments}
My thanks to my collaborators on the D\O\ experiment, particularly the current and past QCD physics group coordinators Mike Strauss, Sabine Lammers, Dmitry Bandurin and Markus Wobisch. All credit for these results goes to this remarkable collaboration, while any errors in the reporting of these results are the author's own.
\end{acknowledgments}

\bigskip 

\end{document}